\def\be{\begin{equation}}
\def\ee{\end{equation}}
\def\ba{\begin{array}}
\def\ea{\end{array}}
\def\1{{\bf 1}}
\def\p{\prime}

\def\t{\tau}
\def\R1{{1\!\! 1}}
\def\Rb{{I\!\! R}}
\def\Nb{{I\!\! N}}
\def\Fb{{I\!\! F}}
\def\Cb{\ \hbox{\vrule width 0.6pt height 6pt depth 0pt
		      \hskip -3.5 pt} C}
\documentstyle[12pt]{article}
\begin{document}
\parskip=3pt
\parindent=18pt
\baselineskip=18pt
\setcounter{page}{1}
\centerline{\Large\bf Integrable Stochastic Ladder Models}
\vspace{6ex}
\begin{center}
Sergio Albeverio \footnote{ SFB 256; BiBoS; CERFIM(Locarno); Acc. Arch.;
USI(Mendriso)}~~~~
and~~~~
Shao-Ming Fei \footnote{ Institute of Applied Mathematics, Chinese Academy 
of Science, Beijing}\\
Institut f{\"u}r Angewandte Mathematik, Universit{\"a}t Bonn, D-53115, Bonn.\\
\end{center} 
\vskip 1 true cm
\parindent=18pt
\parskip=6pt
\begin{center}
\begin{minipage}{5in}
\vspace{3ex}
\centerline{\large Abstract}
\vspace{4ex}
A general way to construct ladder models with certain
Lie algebraic or quantum Lie algebraic symmetries is presented.
These symmetric models give rise to series of integrable systems. 
It is shown that corresponding to these $SU(2)$ 
symmetric integrable ladder models
there are exactly solvable stationary discrete-time (resp.
continuous-time) Markov processes with
transition matrices (resp. intensity matrices) having spectra which
coincide with the ones of the corresponding integrable models.

\bigskip
\medskip


\end{minipage}
\end{center}

\newpage

Integrable models have played significant roles
in statistical and condensed matter physics. Some of them
have been obtained and investigated using an algebraic or coordinate
``Bethe Ansatz method" \cite{faddeev,baxter}.
The intrinsic symmetry of these integrable chain models 
plays an essential role in finding complete sets of
eigenstates of the systems.
On the other hand, stochastic models like 
stochastic reaction-diffusion models,
models describing coagulation/decoagulation, 
birth/death processes, pair-creation/pair-annihilation 
of molecules on a chain, have attracted
considerable interest due to their importance in many physical,
chemical and biological processes \cite{evan}.
The theoretical description of stochastic reaction-diffusion systems
is given by the ``master equation'' which describes the time evolution
of the probability distribution function \cite{alca,4s}. This equation
has the form of a heat equation with potential (i.e., a Schr\"odinger 
equation with ``imaginary time").
If an integrable system with open boundary condition can be transformed
into a stochastic reaction-diffusion system, e.g., by a unitary
transformation between their respective Hamiltonians, looked upon
as self-adjoint operators acting in the respective Hilbert spaces,
then the stochastic model so obtained is exactly solvable with the same
energy spectrum as the one of the integrable system
\cite{alca,ritten,henkel}.

In \cite{intmod}, we have
presented a general procedure to construct open chain models having
a certain Lie algebra or quantum Lie algebra symmetry
by using the coproduct properties of bi-algebras. These models can
be reduced to integrable ones via a detailed representation of the
symmetry algebras involved. In recent years spin ladders
have attracted considerable attention, due to the
developing experimental results on ladder materials and
the hope to get some insight into the physics of metal-oxide
superconductors \cite{DagottoRice96}.
In this letter we study the
construction of ladder models with certain Lie algebra or quantum Lie
algebra symmetry. We show that the integrable quantum spin ladder model
discussed in \cite{ladder} can be obtained in this way and it can be transformed 
into both stationary discrete-time (discrete reaction-diffusion models)
and stationary continuous-time Markov
processes with transition matrices resp. intensity matrices
having the same spectra as the ones of
this $SU(2)$ invariant integrable ladder model.

Let $A$ be an associative Lie bi-algebra 
with basis $e=\{e_\alpha\}$,
$\alpha=1,2,...,n$, satisfying the Lie commutation relations
$[e_\alpha,e_\beta]=C_{\alpha\beta}^\gamma e_\gamma$,
where $C_{\alpha\beta}^\gamma$ are the structure constants with respect
to the base $e$.
Let $\Delta$ (resp. $C(e)$) be the coproduct operator (resp. Casimir
operator) of the algebra $A$.
The coproduct operator action on the Lie algebra elements is given by
$\Delta e_\alpha=e_\alpha\otimes\1+\1\otimes e_\alpha$,
$\1$ stands for the identity operator. It can be immediately checked that
$[\Delta e_\alpha,\Delta e_\beta]=C_{\alpha\beta}^\gamma \Delta e_\gamma$ 
and $[\Delta C(e),\Delta e_\alpha]=0,~~~\alpha=1,2,...,n$.

Let us consider a two-leg ladder with $L$ rungs.
To each point at the $i$-th rung, $i=1,...,L$, and $\theta$-th leg, $\theta =1,2$,
of the ladder we associate a (finite dimensional complex) Hilbert space $H_i^\theta$. 
We can then associate to the whole ladder
the tensor product $H_1^1 \otimes H_1^2 \otimes H_2^1 \otimes H_2^2\otimes...
\otimes H_{L}^1 \otimes H_{L}^2$.
The generators of the algebra $A$ acting on this Hilbert space
associated with the above ladder are given by
$E_\alpha=\Delta^{2L-1} e_\alpha$, $\alpha=1,2,...,n$,
where we have defined
\be\label{deltam}
\Delta^m=(\underbrace{{\1}\otimes ... \otimes{\1}}_{m~times}\otimes\Delta)
...({\1}\otimes{\1}\otimes\Delta)
({\1}\otimes\Delta)\Delta,~~\forall\, m\in\Nb.
\ee
$E_\alpha$ also generates the Lie algebra $A$:
$[E_\alpha,E_\beta]=C_{\alpha\beta}^\gamma E_\gamma$.

Let
\be\label{h}
h=\sum_{i=1}^3 \sum_{j=1}^2 a_{ij}\Delta_i^2\Delta_j^1\Delta C(e),
\ee
where $\Delta_1^1=(\Delta\otimes\1)$, $\Delta_2^1=(\1\otimes\Delta)$,
$\Delta_1^2=(\Delta\otimes\1\otimes\1)$, $\Delta_2^2=(\1\otimes\Delta\otimes\1)$,
$\Delta_3^2=(\1\otimes\1\otimes\Delta)$, $a_{ij}\in\Cb$ such that $h$ is hermitian.
Let $\Fb$ denote a real entire function defined on the $2L$-th tensor space 
$A\otimes A\otimes...\otimes A$ of the algebra $A$. We call 
\be\label{18}
H=\sum_{i=1}^{L-1} \Fb(h)_{i,i+1}
\ee
the (quantum mechanics) Hamiltonian associated with the ladder.
Here $\Fb(h)_{i,i+1}$ means that the four-fold tensor
element $\Fb(h)$ is associated with the $i$ and $i+1$-th rungs of the
ladder and acts on the space $H_i^1 \otimes H_i^2 \otimes H_{i+1}^1 
\otimes H_{i+1}^2$, i.e.,
\be\label{pp}
\Fb (h)_{i,i+1}=\1_1^1 \otimes \1_1^2\otimes...\otimes\1_{i-1}^1 \otimes \1_{i-1}^2 \otimes \Fb(h)
\otimes\1_{i+2}^1\otimes \1_{i+2}^2\otimes... \otimes\1_{L}^1 \otimes \1_{L}^2.
\ee

{\sf [Theorem 1]}. The Hamiltonian $H$ is a self-adjoint operator acting
in $H_1^1 \otimes H_1^2 \otimes H_2^1 \otimes H_2^2\otimes...
\otimes H_{L}^1 \otimes H_{L}^2$ and is invariant under the algebra $A$.
	
{[\sf Proof].} That $H$ is self-adjoint is immediate from the definition.
To prove the invariance of $H$
it suffices to prove $[H,E_\alpha]=0$, $\alpha=1,2,...,n$.

From the formula for the above coproduct we have
\be\label{19}
E_\alpha=\sum_{i=1}^L(e_{\alpha})_i,
\ee
where $(e_{\alpha})_i=\1_1^1\otimes\1_1^2\otimes...\otimes\1_{i-1}^1\otimes\1_{i-1}^2\otimes
(e_{\alpha}\otimes\1_i^2\otimes+
\1_i^1\otimes e_{\alpha})\otimes
\1_{i+1}^1\otimes\1_{i+1}^2\otimes... \otimes\1_{L}^1\otimes\1_{L}^2$. 

From $[\Delta C(e),\Delta e_\alpha]=0$, $\alpha=1,2,...,n$, it follows easily that
$[h,\Delta^2 e_{\alpha}]=0$, where $\Delta^2 e_{\alpha}=(\1\otimes\1\otimes\Delta)
(\1\otimes\Delta)(\Delta)e_{\alpha}$, as defined in (\ref{deltam}).
Obviously $[\Fb(h)_{i,i+1},(e_{\alpha})_j]=0$, $\forall j\neq i,i+1$.
Therefore we have, for all
$\alpha=1,2,...,n$:
\be\label{he}
\ba{rcl}
[H,E_\alpha]&=&
\left[\displaystyle\sum_{i=1}^{L-1} \Fb(h)_{i,i+1},
\displaystyle\sum_{j=1}^{L-1}(e_{\alpha})_j\right]\\[5mm]
&=&\displaystyle\sum_{i=1}^{L-1}\left[\Fb(h)_{i,i+1},
\displaystyle\sum_{j=1}^{i-1}(e_{\alpha})_j+
\displaystyle\sum_{k=i+2}^{L}(e_{\alpha})_k
+(e_{\alpha})_i+(e_{\alpha})_{i+1}\right]\\[5mm]
&=&\displaystyle\sum_{i=1}^{L-1}\left[\Fb(h)_{i,i+1},
(e_{\alpha})_i+(e_{\alpha})_{i+1}\right]
=\displaystyle\sum_{i=1}^{L-1}\left[
\Fb(h)_{i,i+1},(\Delta^2 e_\alpha)_{i,i+1}\right]=0.
\ea
\ee
\hfill $\rule{3mm}{3mm}$

Let $V$ be a complex vector space and $\check{R}$ be the solution of 
quantum Yang-Baxter equation (QYBE) \cite{baxter,yang} without
spectral parameters, see e.g. \cite{pressley}. Then $\check{R}$ 
takes values in $End_{\Cb}(V\otimes V)$. The QYBE is
\be\label{37}
\check{R}_{12}\check{R}_{23}\check{R}_{12}=
\check{R}_{23}\check{R}_{12}\check{R}_{23},
\ee
where  $\check{R}_{12}=\check{R}\otimes id$, $\check{R}_{23}=id \otimes
\check{R}$ and $id$ is the identity operator on $V$. 

In the following we say that a ladder model
having a (quantum mechanical) Hamiltonian of the form
\be\label{38}
H=\sum_{i=1}^{L-1}({\cal H})_{i,i+1}
\ee
is integrable in the sense that the operator
${\cal H}$ satisfies the QYBE relation (\ref{37}), i.e.,
\be\label{39}
({\cal H})_{12}({\cal H})_{23}({\cal H})_{12}=
({\cal H})_{23}({\cal H})_{12}({\cal H})_{23},
\ee
where $({\cal H})_{12}={\cal H}\otimes id$ and $({\cal H})_{23}=id
\otimes {\cal H}$. ${\cal H}$ is a solution of the Yang-Baxter
equation without spectral parameters. Correspondingly the $i$-th
complex vector space $V_i$ now stands for $H_i^1 \otimes H_i^2$.
After Baxterization the
Hamiltonian system (\ref{38}) satisfying relation (\ref{39}) can in
principle be exactly solved by the algebraic Bethe Ansatz method, see e.g.
\cite{faddeev}.

We consider ladder models with $SU(2)$ symmetry. Let $S_i$, $i=1,2,3$, and $C$ 
be the generators of the algebra $SU(2)$ and Casimir operator respectively.
The coproduct of the algebra is given by $\Delta S_i=\1\otimes S_i +S_i\otimes \1$,
$i=1,2,3$. Taking into account that $\Delta_i^j \Fb (e)=\Fb(\Delta_i^j e)$, $i=1,2,3$,
$j=1,2$, $\forall\, e \in SU(2)$, the generic $h$ is of the form
$\Fb(C_1,C_2,C_3)$, where 
$$
\ba{l}
C_1=\displaystyle\sum_{i=1}^{3}(
S_i \otimes \1 \otimes \1 \otimes S_i+ \1 \otimes S_i \otimes \1 \otimes S_i 
+\1 \otimes \1 \otimes S_i \otimes S_i),\\
C_2=\displaystyle\sum_{i=1}^{3}(
S_i \otimes \1 \otimes \1 \otimes S_i+ S_i \otimes S_i \otimes \1 \otimes \1 
+S_i \otimes \1 \otimes S_i \otimes \1),\\
C_3=\displaystyle\sum_{i=1}^{3}(
S_i \otimes \1 \otimes \1 \otimes S_i+ \1 \otimes S_i \otimes \1 \otimes S_i 
+S_i \otimes \1 \otimes S_i \otimes \1 + \1 \otimes S_i \otimes S_i \otimes \1).
\ea
$$ 

In the spin-$\frac{1}{2}$ representation of the algebra $SU(2)$, the solutions of
the QYBE (\ref{39}) are $16\times 16$ matrices. For instance, it is easy to check that
\be\ba{rcl}
{\cal H}_0&=&\frac{ 108d - 55f}{108} C_{111} 
+\frac{ -72 d + 104 f} {288} C_{112} 
+\frac{ -486 d + 211 f}{270} C_{113} + \frac{ -756 d + 370 f} {216}C_{121}\\[4mm]
&& 
- \frac{29 f}{108} C_{122} 
+ \frac{ 90 d - 31 f}{36} C_{123} + \frac{ 2 d - f}{2} C_{131} 
+ \frac{ -54 d + 26 f}{108}C_{132}\\[4mm]
&& + \frac{ -108 d + 43 f}{540}C_{133}
+ \frac{ -216 d + 80 f}{864}C_{211} 
+ \frac{11 f}{108}C_{212} 
+\frac{ 216 d - 119 f}{108}C_{213}
\ea
\ee
satisfies (\ref{39}) for all $d,f\in\Rb$, 
where $C_{ijk}\equiv C_i\cdot C_j\cdot C_k$, $i,j,k=1,2,3$.

The corresponding solution related to the 
$SU(2)$-symmetric integrable ladder model in \cite{ladder}
can be expressed as
\be\label{rr0}
\ba{rcl}
{\cal H}&=&-\frac{5}{48}C_{111}
-  \frac{11}{32}C_{112} 
- \frac{61}{30}C_{113} 
 - \frac{41}{48}(C_{121}-C_{122}) + 
\frac{21}{16}C_{123}\\[4mm] 
	  &&+ \frac{3}{4}C_{131} 
-\frac{17}{12}C_{132}+\frac{173}{240}C_{133}
+ \frac{55}{96}C_{211} 
- \frac{5}{3}C_{212} + \frac{131}{48}C_{213}.
\ea
\ee
Through baxterization, ${\cal H}(x)=(x-1){\cal H}+16\, {\rm I}_{16\times 16}$
satisfies the QYBE with spectral parameters:
${\cal H}_{12}(x){\cal H}_{23}(x y){\cal H}_{12}(y)
={\cal H}_{23}(y){\cal H}_{12}(x y){\cal H}_{23}(x)$, where
${\cal H}_{12}(\cdot)={\cal H}(\cdot)\otimes {\rm I}_{4\times 4}$,
${\cal H}_{23}(\cdot)={\rm I}_{4\times 4}\otimes {\cal H}(\cdot)$,
${\rm I}_{n\times n}$ denotes the $n\times n$ identity matrix.
The model can be exactly solved using a algebraic Bethe Ansatz method.
It describes a periodic spin ladder system with both isotropic
exchange interactions and biquadratic interactions:
$$
\begin{array}{rcl}
H&=&\displaystyle\frac12
\sum_{i=1}^{L-1}(\frac12+2{\bf S}_{1,i}\cdot{\bf S}_{1,i+1})
(\frac12+2{\bf S}_{2,i}\cdot{\bf S}_{2,i+1})
-\displaystyle\frac12\sum_{i=1}^{L-1}(\frac12+2{\bf S}_{1,i}\cdot{\bf S}_{2,i+1})
(\frac12+2{\bf S}_{2,i}\cdot{\bf S}_{1,i+1})\\[4mm]
&&+\displaystyle\frac{5}{6}\sum_{i=1}^{L-1}
(\frac12+2{\bf S}_{1,i}\cdot{\bf S}_{2,i})
(\frac12+2{\bf S}_{1,i+1}\cdot{\bf S}_{2,i+1}),
\end{array}
$$
where ${\bf S}_{\theta,i}=(\sigma^x_{\theta,i},\sigma^y_{\theta,i},\sigma^z_{\theta,i})/2$, 
$\sigma^x,\sigma^y,\sigma^z$ are Pauli matrices.
${\bf S}_{1,i}$ (resp. ${\bf S}_{2,i}$) is the spin operator on
the first (resp. second) leg of the $i$-th rung of the ladder.

It is further shown that for a more general form of (\ref{rr0}),
\be\label{rr}
\ba{rcl}
{\cal H}^\prime&=&\frac{ -45 + 23\,a - 4\,b - 28\,c }{432}
      C_{111} + \frac{ -99 - 3\,a - 3\,b - c }{288}
      C_{112} +  \frac{ -1098 - 91\,a - 118\,b - 16\,c }{540} C_{113}\\[4mm] 
&& + \frac{ -369 - 97\,a - 70\,b + 50\,c }{432} C_{121} + 
\frac{ 396 + 4\,a + 31\,b + 25\,c }{432} C_{122} + 
\frac{ 189 + 29\,a + 20\,b - 4\,c }{144} C_{123}\\[4mm] 
	  &&+ \frac{3}{4}C_{131} + \frac{ -306 - 2\,a - 29\,b - 14\,c }{216}
      C_{132} + \frac{ 1557 - 71\,a + 172\,b + 124\,c }{2160}C_{133}\\[4mm]
	  && + \frac{ 495 - a + 53\,b + 47\,c }{864} C_{211} + 
\frac{ -720 - 22\,a - 49\,b - 43\,c }{432} C_{212} + 
\frac{ 1179 + 91\,a + 118\,b + 16\,c }{432} C_{213}
\ea
\ee
with $a,b,c\in\Rb$, the corresponding ladder model $H^\prime=\displaystyle
\sum_{i=1}^{L-1}{\cal H}^\prime_{i,i+1}$ can also be exactly solved by a
coordinate Bethe ansatz \cite{ladder}.

We consider now stochastic processes \cite{markov} on a ladder.
Let $(\Omega,P)$ be a probability space, with
$\Omega$ the finite sample space and $P$ the probability
measure defined on the $\sigma$-algebra of all subsets of
$\Omega$. For a discrete time stationary Markov chain $\{X_i\}$, $i\in\Nb$, with
underlying probability space $(\Omega,P)$ and a finite
state space $S=\{1,2,3,...,m\}$, there are $m^2$ transition
probabilities $\{p_{\alpha\beta}\}$, $\alpha,\beta=1,2,...,m$.
The stochastic
transition matrix $P=(p_{\alpha\beta})$ has the following properties:
\be\label{pm}
p_{\alpha\beta}\ge 0,~~~\sum_{\alpha=1}^m p_{\alpha\beta}=1,~~~\alpha,\beta=1,2,...,m.
\ee

For a stationary continuous-time real-valued stochastic process,
$\{X_t\}_{t\in\Rb_+}$ (on the probability space $(\Omega,P)$),
the transition semigroup $P(t)=P(X_{t=j}\vert X_0=i)$
is generated by an intensity matrix $Q=(q_{\alpha\beta})$ with the properties:
\be\label{qm}
q_{\alpha\beta}\ge 0,~~~\alpha\neq\beta,~~~~q_{\alpha\alpha}
=-\sum_{\alpha\neq\beta}q_{\alpha\beta},~~~\alpha,\beta=1,2,...,m.
\ee

The transition matrix $P$ (resp. intensity matrix $Q$) defines
the stochastic processes on a ladder. In the following we call
a ladder associated with above stochastic processes, for instance,
particles jumping randomly on the ladder, characterized by the matrices
$P$ and $Q$ a Markov ladder, though geometrically it is equivalent to 
a chain with particular non-nearest neighbor interactions.
If the eigenvalues and eigenstates of $P$ resp. $Q$ are
known, then exact results concerning the stochastic processes, such as
time-dependent averages and correlations, can be obtained. We say that
a Markov ladder is integrable (resp. $SU(2)$-symmetric) if the 
eigenvalues and eigenstates of the related transition matrix $P$ or 
intensity matrix $Q$ can be exactly solved (resp. is $SU(2)$ invariant).
 
To every site on the $i$-th rung and $\theta$-th leg of the ladder we associate
states described by the variable $\t_i^j$ taking values $0$ and $1$
(conventionally a vacancy at the site is associated with the state $0$ and an occupied
state is associated with the state $1$).
The state space of this algebraic ladder 
is then finite and has a total of $m=2^{2L}$ states.

{\sf [Theorem 2]}. The following matrix
\be\label{pan}
P_{SU(2)}=\displaystyle\frac{1}{4(L-1)(18 + 4a + 4b + c)}
\sum_{i=1}^{L-1}{\cal H}^{\prime\prime}_{i,i+1},
\ee
defines a stationary discrete-time $SU(2)$-symmetric integrable Markov ladder for
$a$, $b$, $c \geq 0$, $a+2b-16 \geq 0$.
The operator ${\cal H}^{\prime\prime}$ is given by
\be\label{hpp}
{\cal H}^{\prime\prime}=
\left(
\ba{cccccccccccccccc}
  a_1& a_2& a_2& a_2& a_3& a_4& a_4& a_4& a_3& a_4& a_4& a_4& a_3& a_4& a_4& a_4\\[3mm] 
  a_2& a_5& a_6& a_6& a_7& a_3& a_8& a_8& a_8& a_9& a_4& a_4& a_8& a_9& a_4& a_4\\[3mm] 
  a_2& a_6& a_5& a_6& a_8& a_4& a_9& a_4& a_7& a_8& a_3& a_8& a_8& a_4& a_9& a_4\\[3mm] 
  a_2& a_6& a_6& a_5& a_8& a_4& a_4& a_9& a_8& a_4& a_4& a_9& a_7& a_8& a_8& a_3\\[3mm] 
  a_3& a_7& a_8& a_8& a_5& a_2& a_6& a_6& a_9& a_8& a_4& a_4& a_9& a_8& a_4& a_4\\[3mm] 
  a_4& a_3& a_4& a_4& a_2& a_1& a_2& a_2& a_4& a_3& a_4& a_4& a_4& a_3& a_4& a_4\\[3mm] 
  a_4& a_8& a_9& a_4& a_6& a_2& a_5& a_6& a_8& a_7& a_3& a_8& a_4& a_8& a_9& a_4\\[3mm] 
  a_4& a_8& a_4& a_9& a_6& a_2& a_6& a_5& a_4& a_8& a_4& a_9& a_8& a_7& a_8& a_3\\[3mm] 
  a_3& a_8& a_7& a_8& a_9& a_4& a_8& a_4& a_5& a_6& a_2& a_6& a_9& a_4& a_8& a_4\\[3mm] 
  a_4& a_9& a_8& a_4& a_8& a_3& a_7& a_8& a_6& a_5& a_2& a_6& a_4& a_9& a_8& a_4\\[3mm] 
  a_4& a_4& a_3& a_4& a_4& a_4& a_3& a_4& a_2& a_2& a_1& a_2& a_4& a_4& a_3& a_4\\[3mm] 
  a_4& a_4& a_8& a_9& a_4& a_4& a_8& a_9& a_6& a_6& a_2& a_5& a_8& a_8& a_7& a_3\\[3mm] 
  a_3& a_8& a_8& a_7& a_9& a_4& a_4& a_8& a_9& a_4& a_4& a_8& a_5& a_6& a_6& a_2\\[3mm] 
  a_4& a_9& a_4& a_8& a_8& a_3& a_8& a_7& a_4& a_9& a_4& a_8& a_6& a_5& a_6& a_2\\[3mm] 
  a_4& a_4& a_9& a_8& a_4& a_4& a_9& a_8& a_8& a_8& a_3& a_7& a_6& a_6& a_5& a_2\\[3mm] 
  a_4& a_4& a_4& a_3& a_4& a_4& a_4& a_3& a_4& a_4& a_4& a_3& a_2& a_2& a_2& a_1
\ea
\right)
\ee
where $a_1=66 + a + 4b + 4c$, $a_2=-10 + a + 2b$, $a_3=6 + a + 2b$,
$a_4=2 + a$, $a_5=54 + a + 4b + 4c$, $a_6=-16 + a + 2b$,
$a_7=14 + a$, $a_8=8 + a$, $a_9=a + 2b$. ${\cal H}^{\prime\prime}_{i,i+1}$
acts on the $i$ and $i+1$ rungs as defined in (\ref{pp}).

{[\sf Proof].} For the integrable ladder model with Hamiltonian 
$H^\prime=\displaystyle\sum_{i=1}^{L-1}{\cal H}^{\prime}_{i,i+1}$, 
the system  remains integrable
if one adds to $H^\prime$ a constant term and multiplies $H^\prime$ by a constant
factor. Moreover the spectrum of $H^\prime$
will not be changed if one changes the local basis of the rungs, i.e., the following
Hamiltonian $H^{\p\p}$, defined by
\be\label{hp}
H^{\p\p}=BH^\prime B^{-1},~~~~B=\displaystyle\otimes_{i=1}^{L}B_i,
\ee
where $B_i$ are $4\times 4$ non singular matrices, 
has the same eigenvalues as $H^\prime$.

It is straightforward to prove that ${\cal H}^{\prime\prime}=B {\cal H}^{\prime}B^{-1}$, where
$$
B=\left(
\ba{cccc}
-1 & 1 & 0 & 0\\[3mm]
1 & 1/2 & -1/2 & 1\\[3mm]
0 & -1/2 & -3/2 & 0\\[3mm]
0 & 1 & 0 & -1
\ea
\right).
$$
Therefore the Hamiltonian systems $H^{\p}$ and 
$H^{\p\p}=\displaystyle
\sum_{i=1}^{L-1}{\cal H}^{\prime\prime}_{i,i+1}$ satisfy the relation (\ref{hp})
with $B_i=B$, $i=1,2,...,L$. Hence $H^{\p\p}$ is also $SU(2)$-symmetric and integrable
with the same spectrum as $H^{\p}$.

For $a+2b \geq 0$, as the entries of ${\cal H}^{\prime\prime}$ are positive,
$H^{\prime\prime}_{\alpha\beta}\geq 0$, $\alpha,\beta=1,2,...,2^{2L}$. From (\ref{hpp}) we
also have $\sum_{\alpha=1}^{16}{\cal H}^{\p\p}_{\alpha\beta}=4(18+4a+4b+c)$, $\forall\, \beta=
1,2,...,16$. By the definition (\ref{pm}) $P_{SU(2)}$ is the transition matrix
of a stationary discrete-time $SU(2)$-symmetric integrable Markov ladder.
\hfill $\rule{3mm}{3mm}$

The state space of this Markov processes associated with the 
stochastic matrix $P_{SU(2)}$ is $S=(1,2,...,2^{2L})$.
Generally there is no closed subset $C$ of
the state space $S$ such that $(P_{SU(2)})_{ij}=0$ for all $i\in C$ and
$j\not\in C$. In a certain parameter region of the $a,b,c$
there are nonempty closed sets other than $S$ itself and the Markov
ladder becomes reducible. However there exists no
absorbing state in this Markov ladder.

By using results in the proof of theorem 2, we have also
the following integrable stationary continuous-time Markov ladder:

{\sf [Theorem 3]}. The matrix 
\be\label{qan}
Q_{SU(2)}=H^{\p\p}-4(L-1)(18 + 4a + 4b + c)=
\sum_{i=1}^{L-1}({\cal H}^{\p\p}-4(18 + 4a + 4b + c))_{i,i+1}
\ee
is the intensity matrix of a stationary continuous-time Markov ladder.

We have discussed the construction of integrable 
ladder models with Lie algebra symmetry.
It is shown that the stochastic processes
correspond to the $SU(2)$ symmetric integrable ladder 
models define exactly solvable stationary discrete-time (resp.
continuous-time) Markov ladder with
transition matrices (resp. intensity matrices) which coincide with those
of the corresponding integrable models.

Integrable ladder models with quantum algebraic
symmetry and the related Markov processes can be investigated in a
similar way. Let $e=\{e_\alpha,f_\alpha,h_\alpha\}$, $\alpha=1,2,...,n$, 
be the Chevalley basis of a Lie algebra $A$ with rank $n$.
Let $e^\p=\{e_\alpha^\p,f_\alpha^\p,h_\alpha^\p\}$, 
$\alpha=1,2,...,n$, be the corresponding elements of the
quantum (q-deformed) Lie algebra  $A_q$. We denote by $r_\alpha$ the simple
roots of the Lie algebra $A$. The Cartan matrix $(a_{\alpha\beta})$ is then
\be\label{22}
a_{\alpha\beta}=\frac{1}{d_\alpha}(r_\alpha\cdot r_\beta),~~~
d_\alpha=\frac{1}{2}(r_\alpha\cdot r_\alpha).
\ee
The coproduct operator $\Delta^\prime$ of the quantum algebra $A_q$ is
given by
\begin{eqnarray}
\Delta^\p h_\alpha^\p&=&h_\alpha^\p\otimes\1+\1\otimes h_\alpha^\p,\label{25}\\[3mm]
\Delta^\p e_\alpha^\p&=&e_\alpha^\p\otimes q^{-d_\alpha h_\alpha^\p}
+q^{d_\alpha h_\alpha^\p}\otimes e_\alpha^\p,\label{26}\\[3mm]
\Delta^\p f_\alpha^\p&=&f_\alpha^\p\otimes q^{-d_\alpha h_\alpha^\p}
+q^{d_\alpha h_\alpha^\p}\otimes f_\alpha^\p,\label{27}
\end{eqnarray}
$q\in\Cb$, $q^{d_\alpha}\neq\pm1,0$. In the following we use the notations
$\Delta^{\p\,m}$ and $\Delta_j^{\p\,i}$ defined similarly as in (\ref{deltam})
and (\ref{h}).

{\sf [Theorem 4]}. The ladder model defined by the following Hamiltonian
acting in $H_1^1 \otimes H_1^2 \otimes H_2^1 \otimes H_2^2\otimes...
\otimes H_{L}^1 \otimes H_{L}^2$ is invariant under the quantum algebra $A_q$:
\be\label{33}
H_q=\sum_{i=1}^{L-1}\Fb(h_q)_{i,i+1},
\ee
where $h_q=\sum_{i=1}^3 \sum_{j=1}^2 a_{ij}\Delta_i^{\p\, 2}\Delta_j^{\p\,1}\Delta C_q(e^\prime)$,
$C_q(e^\prime)$ is the Casimir operator of $A_q$.

{\sf [Proof]}. The generators of $A_q$ on the ladder are given by
\be\label{32}
\ba{rcl}
H^\p_\alpha&=&\Delta^{\p\, 2L-1} h^\p_\alpha=
\displaystyle\sum_{i=1}^{L}
\1_1^1\otimes\1_1^2\otimes...
(h^\p_\alpha \otimes\1_i^2\otimes+\1_i^1\otimes h^\p_\alpha)
\otimes... \otimes\1_{L}^1\otimes\1_{L}^2\\[4mm]
E^\p_\alpha&=&\displaystyle\sum_{i=1}^{L}
q^{d_\alpha h^\p_\alpha}\otimes...
\otimes(e^\p_\alpha \otimes\1_i^2\otimes+\1_i^1\otimes e^\p_\alpha)\otimes...
\otimes q^{-d_\alpha h^\p_\alpha},\\[6mm]
F^\p_\alpha&=&\displaystyle\sum_{i=1}^{L}
q^{d_\alpha h^\p_\alpha}\otimes...
\otimes (f^\p_\alpha \otimes\1_i^2\otimes+\1_i^1\otimes f^\p_\alpha)\otimes...
\otimes q^{-d_\alpha h^\p_\alpha}.
\ea
\ee

From
$[\Delta^\p \Fb(C_q(e^\prime)), \Delta^\p a]=0,~\forall a\in A_q$
and $\Delta^\p q^{\pm d_\alpha h_\alpha^\p}=q^{\pm d_\alpha h_\alpha^\p}\otimes 
q^{\pm d_\alpha h_\alpha^\p}$, we have
$[h_q,\Delta^{\p\,2}h_\alpha^\p]=[h_q,\Delta^{\p\,2}e_\alpha^\p]=[h_q,\Delta^{\p\,2}f_\alpha^\p]=0$.
Therefore
$$
\ba{rcl}
[H_q,E^\p_\alpha]
&=&\displaystyle\sum_{i=1}^{L-1}\left[
\Fb(h_q)_{i,i+1},(e^\p_\alpha)_i^1\otimes (q^{-d_\alpha h^\p_\alpha})_{i}^2
\otimes (q^{-d_\alpha h^\p_\alpha})_{i+1}^1\otimes (q^{-d_\alpha h^\p_\alpha})_{i+1}^2\right.\\[4mm]
&&+(q^{d_\alpha h^\p_\alpha})_{i}^1
\otimes (e^\p_\alpha)_i^2\otimes (q^{-d_\alpha h^\p_\alpha})_{i+1}^1
\otimes (q^{-d_\alpha h^\p_\alpha})_{i+1}^2\\[4mm]
&&+(q^{d_\alpha h^\p_\alpha})_{i}^1
\otimes (q^{d_\alpha h^\p_\alpha})_{i}^2\otimes (e^\p_\alpha)_{i+1}^1
\otimes(q^{-d_\alpha h^\p_\alpha})_{i+1}^2\\[4mm]
&&\left. +(q^{d_\alpha h^\p_\alpha})_{i}^1
\otimes (q^{d_\alpha h^\p_\alpha})_{i}^2\otimes (q^{d_\alpha h^\p_\alpha})_{i+1}^1
\otimes (e^\p_\alpha)_{i+1}^2\right]\\[4mm]
&=&\displaystyle\sum_{i=1}^{L-1}\left[\Fb(h_q),\Delta^{\p\,2} (e^\p_\alpha)\right]_{i,i+1}=0.
\ea
$$
$[H_q,F^\p_\alpha]=0$ is obtained similarly.
$[H_q,H^\p_\alpha]=0$ can be proved like (\ref{he}).
Hence $H_q$ commutes with the generators of $A_q$ for the ladder.
\hfill $\rule{3mm}{3mm}$

The Hamiltonian system (\ref{33}) is expressed by the quantum
algebraic generators $e^\p=(h^\p_\alpha,e^\p_\alpha,f^\p_\alpha)$. 
Assume now that $e\to e^\prime(e)$ is an
algebraic map from $A$ to $A_q$ (we remark that for rank one algebras, 
both classical and quantum
algebraic maps can be discussed in terms of the two dimensional
manifolds related to the algebras, see  \cite{fa}). We then have
\be\label{35}
H_q=\sum_{i=1}^{L-1} \Fb(h_q(e^\p(e))_{i,i+1}.
\ee
In this way we obtain ladder models having quantum algebraic symmetry
but expressed in terms of the usual Lie algebraic generators $\{e_{\alpha}\}$
with manifest physical meaning.

\end{document}